\documentclass[pre,twocolumn,showkeys,preprintnumbers,amsmath,amssymb,groupedaddress,superscriptaddress]
{revtex4}

\usepackage{graphicx}
\usepackage{dcolumn}
\usepackage{bm}
\usepackage{psfrag}
\usepackage{epsfig}
\usepackage{amsmath}
\usepackage{amssymb}
\usepackage{color}
\usepackage{fancyheadings}
\usepackage{mathbbol}
\usepackage{bbm}
\usepackage{mathrsfs}

\newcommand{\bra}{\langle}
\newcommand{\ve}{\vert}
\newcommand{\ket}{\rangle}

\begin{document}


\title{Spectral densities and partition functions of modular quantum systems as derived from a central limit theorem}

\author{Michael Hartmann} 
\email{michael.hartmann@dlr.de}
\affiliation{Institute of Thechnical Physics, DLR Stuttgart}
\affiliation{Institute of Theoretical Physics I, University of Stuttgart}
\author{G\"unter Mahler}
\affiliation{Institute of Theoretical Physics I, University of Stuttgart}
\author{Ortwin Hess}
\affiliation{Advanced Technology Institute, University of Surrey}

\date{\today}

\begin{abstract}
Using a central limit theorem for arrays of interacting quantum systems, we give analytical expressions
for the density of states and the partition function at finite temperature of such a system,
which are valid in the limit
of infinite number of subsystems. Even for only small numbers of subsystems we find good accordance with
some known, exact results.
\end{abstract}

\maketitle

%
%
\section{Introduction}

In order to study a quantum system in full detail, its Hamiltonian needs to be diagonalized.
With increasing dimension of the Hilbert space, the diagonalization of an operator becomes a very tedious task.
For lattices or arrays of interacting subsystems or particles, the dimension of the Hilbert space scales
as $m^n$, where $m$ is the dimension of the Hilbert space of one particle and $n$ is the number of particles.
Thus, apart from some exceptions \cite{Korepin1993}, it is even numerically impossible to exactly
determine the eigenvalues of those models.

Fortunately, considerable understanding can be obtained already from functions of the eigenvalues
without knowing each individual eigenvalue. For example all thermodynamical quantities of a system are
determined by its partition function \cite{Toda1992,Kubo1985,Juettner1996}.

In this paper, we present a novel approach, that allows to derive analytical expressions for all quantities
whose operator is a function of the Hamiltonian for chains or lattices of very many interacting particles
\cite{Lieb1966,Mahler1998,Jaksch1998}.

Our approach is based on a central limit theorem \cite{Linnik1971, Billingsley1995}
for quantum systems with nearest neighbor interactions.
It is also valid if each subsystem does not only interact with its nearest neighbors but with
a fixed, finite number of neighbors.

Systems of that interaction topology play a central role in condensed matter theory
\cite{Kittel1983,Lieb2001}.
Most models, which are currently used to describe strongly interacting electrons, belong to this category.

In the analysis of the thermodynamics of systems of interacting particles, potential phase transitions
are of central interest.
Being based on a central limit theorem, our approach becomes more exact with increasing number of subsystems.
It is precisely this limit of infinite number of subsystems, that is relevant for the study of the
possible phase transitions \cite{Kubo1985, Sachdev1999}.

The paper is organized as follows: In section \ref{model}, we present the class of models we address an
introduce the notations we use. Section \ref{theorem} contains the central limit theorem, our approach is
based on. The proof is not displayed here as it already appeared in a previous publication \cite{Hartmann2003}.
In the following two sections, \ref{zustndsdichte} and \ref{verteilungsf}, we give the analytical
expressions for the density of states and the partition function, which are a straight forward application
of the theorem of section \ref{theorem}. In the next section \ref{numeric}, we numerically evaluate these
expressions for an Ising spin chain and compare our results with the results of an exact diagonalization.
Section \ref{discuss} contains a discussion of some limits and problems of our approach.
Finally section \ref{summary} summarizes the results.
%
%
\section{Model and Notation}
\label{model}

We consider a chain of quantum systems with next neighbor interactions.
The entire system is described by a Hamiltonian $H$ which is a linear, self-adjoint operator on
a separable, complex Hilbert space $\mathscr{H}$.
The Hilbert space $\mathscr{H}$ is a direct product of the Hilbert spaces of the subsystems,
\begin{equation}\label{hilbert}
\mathscr{H} \equiv  \prod_{\mu = 1}^n \otimes \mathscr{H}_{\mu},
\end{equation}
and the Hamiltonian may be written in the form,
\begin{equation}\label{hamil}
H \equiv \sum_{\mu = 1}^n \mathcal{H}_{\mu},
\end{equation}
with
\begin{equation}
\mathcal{H}_{\mu} \equiv \mathbbm{I}^{\otimes \mu-1} \otimes H_{\mu} \otimes \mathbbm{I}^{\otimes n - \mu} \: + \:
\mathbbm{I}^{\otimes \mu-1} \otimes I_{\mu,\mu+1} \otimes \mathbbm{I}^{\otimes n - (\mu + 1)},
\end{equation}
where $H_{\mu}$ is the proper  Hamiltonian of subsystem $\mu$,
and $I_{\mu,\mu+1}$ the interaction of subsystem $\mu$ with subsystem $\mu+1$.
$\mathbbm{I}$ is the identity operator. We chose periodic boundary conditions $I_{n,n+1} = I_{n,1}$.

Let $E_{\varphi}$ be the eigenenergies and, using the Dirac notation \cite{Sakurai1994}, let
$\{ \ve \varphi \ket \}$ be an orthonormal basis of $\mathscr{H}$
consisting of eigenstates of the total system.
\begin{equation}
H \ve \varphi \ket = E_{\varphi} \ve \varphi \ket \enspace \textrm{with} \enspace \enspace
\bra \varphi \ve \varphi' \ket = \delta_{\varphi \varphi'} \, ,
\end{equation}
where $\delta_{\varphi \varphi'}$ is the Kronecker delta.
We denote by $\ve a \ket$ the product states
\begin{equation}\label{prod}
\ve a \ket \equiv \prod_{\mu = 1}^n \otimes \, \ve a_{\mu} \ket,
\end{equation}
built up from some eigenstates $\ve a_{\mu} \ket$ of each subsystem Hamiltonian $H_{\mu}$,
$H_{\mu} \ve a_{\mu} \ket = E_{\mu} \ve a_{\mu} \ket$. Let $E_a$ denote the sum of all $E_{\mu}$,
$E_a = \sum_{\mu=1}^n E_{\mu}$.
The states $\ve a \ket$ are assumed to form an orthonormal basis of $\mathscr{H}$.
We furthermore define,
\begin{eqnarray}
\overline{E}_a & \equiv & \bra a \ve H \ve a \ket, \\
\Delta_a^2 & \equiv & \bra a \ve H^2 \ve a \ket - \bra a \ve H \ve a \ket^2,
\end{eqnarray}
and the operator
\begin{equation}
Z \equiv \frac{H - \overline{E}_a}{\Delta_a}.
\end{equation}
$Z$ has the same eigenstates as $H$,
\begin{equation}
Z \ve \varphi \ket = z_{\varphi} \ve \varphi \ket,
\end{equation}
where the $z_{\varphi}$ denote its eigenvalues.
Note that $H$ and therefore $Z$, $E_{\varphi}$,  $z_{\varphi}$, $\overline{E}_a$ and $\Delta_a$
as well as the two bases $\{ \ve \varphi \ket \}$ and $\{ \ve a \ket \}$ depend on $n$.

The measure of the quantum mechanical distribution of the eigenvalues of $Z$ in the state $\ve a \ket$,
is given by the usual formula,
\begin{equation}
\mathbbm{P}_a \left( z_{\varphi} \in \left[z_1, z_2 \right] \right) = 
\sum_{\{ \ve \varphi \ket : z_1 \le z_{\varphi} \le z_2 \}} \ve \bra a \ve \varphi \ket \ve^2,
\end{equation}
where the sum extends over all states $\ve \varphi \ket$ with eigenvalues in the respective interval.
%
%

\section{Central Limit Theorem}
\label{theorem} 

If the operator $H$ and a state $\ve a \ket$ satisfy
\begin{equation} \label{vacuumfluc}
\Delta_a^2 \ge n \, C
\end{equation}
for all $n$ and some $C > 0$ and if each operator $\mathcal{H}_{\mu}$ is bounded, i.e.
\begin{equation} \label{bounded}
\bra \chi \ve \mathcal{H}_{\mu} \ve \chi \ket \le C'
\end{equation}
for all normalized states $\ve \chi \ket \in \mathscr{H}$ and some constant $C'$,
then the quantum mechanical distribution of the eigenvalues of $Z$ in the state $\ve a \ket$ converges
weakly to a Gaussian normal distribution:
\begin{equation}\label{limitz}
\lim_{n \rightarrow \infty} \mathbbm{P}_a \left( z_{\varphi} \in \left[z_1, z_2 \right] \right) =
\int_{z_1}^{z_2} \frac{\exp \left( - \, z^2 / 2 \right)}{\sqrt{2 \pi}} \, dz
\end{equation}
for all $- \infty < z_1 <z_2 < \infty$.

The rigorous proof of this theorem is given in \cite{Hartmann2003}.
Note that the theorem also holds for lattices of interacting quantum systems in arbitrary dimension.
Furthermore the interaction need not be limited to nearest neighbors. As long as each particle only interacts
with a fixed number of neighbors, the theorem holds.

If each subsystem $H_{\mu}$ has an infinite energy spectrum, condition (\ref{bounded}) is not fulfilled.
The constant $C'$ in (\ref{bounded}) however may be chosen arbitrarily large. A larger $C'$ merely means
that the distributions start to converge at a larger number of subsystems.
Thus condition (\ref{bounded}) requires that the energy of the total system is not concentrated in only
a few subsystems but distributed among the majority of them. Therefore, for subsystems with
an infinite spectrum, the ratio of states where our theorem does not apply is negligible \cite{Hartmann2003}.

For applications in physics, where $n$ is very large but finite, we can use the substitution
\begin{equation}
z = \frac{E - \overline{E}_a}{\Delta_a}
\end{equation}
and write equation (\ref{limitz}) as an integral over energies:
\begin{equation}\label{limit}
\begin{split}
\mathbbm{P}_a \left( E_{\varphi} \in \left[E_1, E_2 \right] \right) =
\hspace{3cm}\\
= \int_{E_1}^{E_2} \frac{1}{\sqrt{2 \pi} \, \Delta_a} \,
\exp \left( - \, \frac{\left(E - \overline{E}_a\right)^2}{2 \, \Delta_a^2} \right) \, dE
\end{split}
\end{equation}
for all $- \infty < E_1 <E_2 < \infty$ in the limit of large $n$.

The equality sign should be understood as an ``asymptotic'' one here.
The same applies to all further equations derived from (\ref{limit}).

Let us finally introduce the following abbreviation for the density associated with the measure
$\mathbbm{P}_a$:
\begin{equation} \label{prob_dens}
w_a(E) \equiv \frac{1}{\sqrt{2 \pi} \, \Delta_a} \,
\exp \left( - \, \frac{\left(E - \overline{E}_a\right)^2}{2 \, \Delta_a^2} \right)
\end{equation}
As a consequence the expectation value of an operator $\mathcal{O}$, which is a function of $H$ and thus
diagonal in the eigenbasis, can be written
\begin{equation} \label{expect_O}
\bra a \ve \mathcal{O} \ve a \ket = \int w_a (E) \, O(E) \, dE
\end{equation}
where $O(E)$ is the eigenvalue of $\mathcal{O}$ belonging to the energy $E$. If $\mathcal{O}$ is not
a function of $H$, degenerate eigenvalues of $H$, where $\mathcal{O}$ takes on different values, are
problematic.

Equation (\ref{limit}) can be used to estimate various quantities of interest in physics.
Among those are spectral densities and partition sums. We consider these two quantities
in the following two sections.

%
%
\section{Spectral densities}
\label{zustndsdichte}

Spectra of energy levels and thus spectral densities are of immense interest in the theory of
quantum systems. They play a central role, e.g. in the analysis of chaotic behavior of their dynamics
\cite{Haake2001,Lages2004}.

For systems, where the above theorem holds, the calculation of spectral densities is straight forward.
Let us first consider the counting function $N(E)$, that is the number of energy levels below a given
threshold energy $E$. It is given by the trace of the operator $\Theta ( E - H )$, where $\Theta$
is the Heaviside step function. Since the trace of
an operator is invariant under basis transformations, we choose to compute it in the basis formed by
the product states $\ve a \ket$;
\begin{equation}\label{countingf}
N(E) = \sum_{\{ \ve a \ket \}} \bra a \ve \Theta ( E - H ) \ve a \ket,
\end{equation}
where the sum extends over all states $\ve a \ket$ of the type (\ref{prod}). 
According to equation (\ref{expect_O}), the expectation value of $\Theta ( E - H )$ in the state $\ve a \ket$
reads
\begin{equation}\label{expecttheta}
\bra a \ve \Theta ( E - H ) \ve a \ket = \int_{E_g}^{E} w_a(E') \, dE',
\end{equation}
where $E_g$ is th energy of the ground state of the system. The density of states $\eta$ is given by the 
derivative of the counting function with respect to the energy, $\eta (E) = dN(E) / dE$;
\begin{equation}\label{statedens}
\eta (E) = \sum_{\{ \ve a \ket \}} w_a (E) = \sum_{\{ \ve a \ket \}}
\frac{e^{ - \frac{\left(E - \overline{E}_a\right)^2}{2 \, \Delta_a^2}}}{\sqrt{2 \pi} \, \Delta_a}
\end{equation}
Since the convergence of the distribution is weak, i.e. only on intervals of nonzero length,
the derivative should be understood according to its definition as a linear approximation
on intervals of arbitrarily small but non vanishing length.
%
%
\section{partition sums}
\label{verteilungsf}

The thermodynamics of a physical system is completely determined by its partition function.
It is therefore of fundamental relevance to know the partition function of the system of interest.
For quantum systems its calculation is extremely demanding since it involves the complete
diagonalization of the Hamiltonian. As the dimension of the Hilbert space grows exponentially with the
number of subsystems, the diagonalization of the Hamiltonian quickly becomes impossible even with
super computers.

Equation (\ref{limit}) allows to give an analytical expression for the partition function at finite
temperatures, which can easily
be evaluated numerically. The partition function is given by the trace of the operator
$\exp (- \beta H)$ with the inverse temperature $\beta$. We again express it in the basis $\{ \ve a \ket \}$:
\begin{equation}\label{partsum}
Z = \sum_{\{ \ve a \ket \}} \bra a \ve e^{- \beta H} \ve a \ket
\end{equation}
The expectation values of $\exp (- \beta H)$ can be computed using equation (\ref{expect_O})
\cite{Hartmann2003a,Hartmann2003b}, they read
\begin{equation}\label{expectZ}
\begin{split}
\bra a \ve e^{- \beta H} \ve a \ket =
\frac{1}{2} \, \exp \left(- \beta \overline{E}_a + \frac{\beta^2 \Delta_a^2}{2} \right) \hspace{1.5cm}\\
\left[\textrm{erfc} \left( \frac{E_g - \overline{E}_a + \beta \Delta_a^2}{\sqrt{2} \, \Delta_a} \right) -
\textrm{erfc} \left( \frac{E_u - \overline{E}_a + \beta \Delta_a^2}{\sqrt{2} \, \Delta_a} \right) \right],
\end{split}
\end{equation}
where $\textrm{erfc} (x)$ is the conjugate Gaussian error function
\cite{Abramowitz1970}.
$E_g$ is the energy of the ground state and $E_u$ the upper limit of the energy spectrum.

Expression (\ref{expectZ}) can be simplified further. The underlying central limit theorem
is valid in the limit of a very large number of subsystems. In that limit, the argument of the second
conjugate error function is always much bigger than the argument of the first. Furthermore, it is always
positive, which makes the second error function term negligible compared to the first \cite{Abramowitz1970}.

Therefore, the partition function $Z$ can be taken to read
\begin{equation}\label{partf}
Z = \sum_{\{ \ve a \ket \}} e^{- \beta \left( \overline{E}_a - E_g \right)} \,
e^{\frac{\beta^2 \Delta_a^2}{2}} \,
\frac{1}{2} \textrm{erfc} \left( \frac{E_g - \overline{E}_a + \beta \Delta_a^2}{\sqrt{2} \, \Delta_a} \right),
\end{equation}
where we have rescaled the energy in the first exponent, so that all appearing energies are positive
and therefore $\exp \left(- \beta \left( \overline{E}_a - E_g \right)\right) \le 1$. The ground state energy $E_g$
in the error function is not a consequence of the rescaling but stems from a cutoff in the integral
(\ref{expect_O}) similar to the one in (\ref{expecttheta}).

In contrast to the expression for the density of states (\ref{statedens}), there appears one quantity
in equation (\ref{partf}) that cannot be obtained without diagonalizing the Hamiltonian.
This is the ground state energy $E_g$. 

The exact value of $E_g$ however is not needed. The cutoff in (\ref{expect_O}) at $E_g$ is only introduced
because we are dealing with a finite number of subsystems. In the limit of infinite number of subsystems,
it is irrelevant since the Gaussian function $w_a(E)$ decays strong enough, so that it becomes negligible
at $E = E_g$. A sufficiently good estimate of $E_g$ can be obtained from the spectral density
(\ref{statedens}).

However, since equation (\ref{limit}) is only an approximation to one term in the sum (\ref{partf}),
errors due to the finite number of subsystems may add up. Nonetheless, the partition function divided by the
number of states may be calculated instead (see section \ref{discuss} for details).

Equations (\ref{statedens}) and (\ref{partf}) are the main result of the present paper.
We now turn to verify their validity for a model that can be treated exactly.

%
%
\section{Numerical verification}
\label{numeric}

In this section we present numerical tests of the two main results (\ref{statedens})
and (\ref{partf}) for an Ising spin chain in a transverse field.
The Hamiltonian of the chain reads
\begin{equation} \label{isinghamil}
H = B \left( - \sum_{i=1}^n \sigma_i^z - K \sum_{i=1}^n \sigma_i^x \otimes \sigma_{i+1}^x \right).
\end{equation}
Here, $\sigma_i^x$ and $\sigma_i^z$ are the Pauli matrices, $2 B$ is the difference between local energy levels
and $K B$ the coupling strength. 

The model (\ref{isinghamil}) can be diagonalized via successive Jordan-Wigner, Fourier and
Bogoliubov transformations \cite{Sachdev1999,Katsura1962}. The eigenvalues of the Hamiltonian
(\ref{isinghamil}) read 
\begin{equation} \label{isinghamild}
E_{\varphi} = \sum_{l=-(n/2)+1}^{n/2} \omega_l \left( n_l (\varphi) - \frac{1}{2} \right)
\end{equation}
where the $n_l (\varphi)$ are fermionic occupation numbers that can take on the two values $0$ and $1$.
The eigenfrequencies $\omega_l$ are given by
\begin{equation} \label{frequency}
\omega_l = 2 B \sqrt{K^2 + 1 - 2 K \cos \left(\frac{2 \pi l}{n}\right)}.
\end{equation}
We chose units where Planck's and Boltzmann's constant are equal to one, $\hbar = k_B = 1$.

For a finite number of spins $n$, a ``density of states'' can be defined with respect to certain energy bins:
We chose the size of the bins to be $B$, so that the density of states $\eta_n (E)$ is defined as
\begin{equation} \label{spec_dens_def}
\eta_n (E) \equiv \frac{\textrm{number of eigenstates with } E_{\varphi} \in \left[E, E + B \right)}{B}.
\end{equation}
The second quantity of interest, the partition function, is given by the standard expression \cite{Sachdev1999}
\begin{equation} \label{partf_def}
Z_n = \prod_{l=-(n/2)+1}^{n/2} \cosh \left( \beta \frac{\omega_l}{2} \right),
\end{equation}
where $\beta = T^{-1}$ is the inverse temperature. Note that the exact quantities, $\eta_n (E)$ and $Z_n$,
carry an index $n$, reflecting the finite number of spins, in contrast to the values of the asymptotic
approximation, $\eta (E)$ and $Z$.

Before we proceed to calculate the density of states and the partition function for the model
(\ref{isinghamil}) with the help of equations (\ref{statedens}) and (\ref{partf}),
let us test whether the central limit theorem (\ref{limit}) is applicable at all, that is
whether conditions (\ref{vacuumfluc}) and (\ref{bounded}) are satisfied.

The energy of each spin is at least $- B$ and at most $B$
so that condition (\ref{bounded}) is fulfilled. The squared width $\Delta_a^2$ reads
\begin{equation}
\Delta_a^2 = n \, B^2 \, K^2,
\end{equation}
where $n$ is the number of spins, and condition (\ref{vacuumfluc}) is also met.
For a large number of spins, the density of states and the partition function of the system at hand
can thus, indeed, be calculated via equations (\ref{statedens}) and (\ref{partf}).

In the model at hand, $\overline{E}_a = E_a$ and $\Delta_a^2 = const$. Therefore, the sum over all states
$\ve a \ket$ can be transformed into a sum over all energies $E_a = k \, 2 \, B - n \, B$,
$k = 0, 1, \dots, n$,
\begin{equation}
\sum_{\{ \ve a \ket \}} = \sum_{k} {n \choose k}.
\end{equation}

\begin{figure}
\psfrag{-15}{\small \raisebox{-0.1cm}{$-15$}}
\psfrag{-10}{\small \raisebox{-0.1cm}{$-10$}}
\psfrag{-5}{\small \raisebox{-0.1cm}{$-5$}}
\psfrag{5}{\small \raisebox{-0.1cm}{$5$}}
\psfrag{10}{\small \raisebox{-0.1cm}{$10$}}
\psfrag{15}{\small \raisebox{-0.1cm}{$15$}}
\psfrag{20}{\small \hspace{+0.2cm} $20$}
\psfrag{40}{\small \hspace{+0.2cm} $40$}
\psfrag{60}{\small \hspace{+0.2cm} $60$}
\psfrag{80}{\small \hspace{+0.2cm} $80$}
\psfrag{100}{\small \hspace{+0.2cm} $100$}
\psfrag{n}{\hspace{-0.4cm} \raisebox{0.1cm}{$\eta \:, \: \eta_n$}}
\psfrag{c1}{$\: E / B$}
\epsfig{file=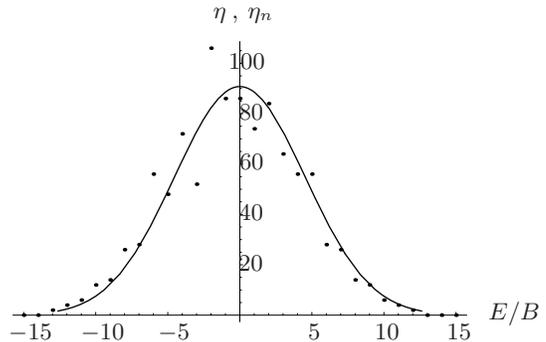,width=7cm}
\caption{Density of states for a chain of 10 spins with $B = K = 1$. The dots show the exact density
$\eta_n$ and the line the approximation $\eta$ ($\eta_n$ and $\eta$ are defined in equations
(\ref{spec_dens_def}) and (\ref{statedens}) respectively).}
\label{density_10_1}
\end{figure}

\begin{figure}
\psfrag{-20}{\small \raisebox{-0.1cm}{$-20$}}
\psfrag{-10}{\small \raisebox{-0.1cm}{$-10$}}
\psfrag{10}{\small \raisebox{-0.1cm}{$10$}}
\psfrag{20}{\small \raisebox{-0.1cm}{$20$}}
\psfrag{500}{\small \hspace{+0.2cm} $500$}
\psfrag{1000}{\small \hspace{+0.2cm} $1000$}
\psfrag{1500}{\small \hspace{+0.2cm} $1500$}
\psfrag{2000}{\small \hspace{+0.2cm} $2000$}
\psfrag{n}{\hspace{-0.4cm} \raisebox{0.1cm}{$\eta \:, \: \eta_n$}}
\psfrag{c1}{$\: E / B$}
\epsfig{file=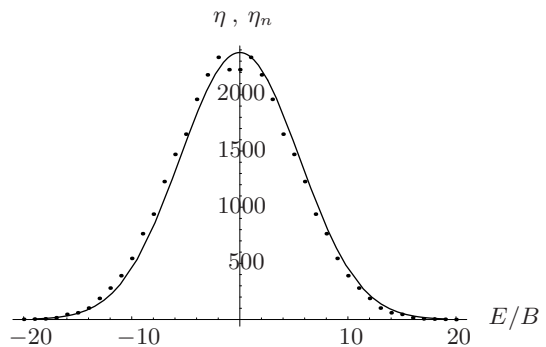,width=7cm}
\caption{Density of states for a chain of 15 spins with $B = K = 1$. The dots show the exact density
$\eta_n$ and the line the approximation $\eta$ ($\eta_n$ and $\eta$ are defined in equations
(\ref{spec_dens_def}) and (\ref{statedens}) respectively).}
\label{density_15_1}
\end{figure}

\begin{figure}
\psfrag{-1.1}{\small \raisebox{-0.15cm}{$10^{-1}$}}
\psfrag{1.1}{\small \raisebox{-0.15cm}{$10^1$}}
\psfrag{2.1}{\small \raisebox{-0.15cm}{$10^2$}}
\psfrag{3.1}{\small \raisebox{-0.15cm}{$10^3$}}
\psfrag{4.1}{\small \raisebox{-0.15cm}{$10^4$}}
\psfrag{5.1}{\small \raisebox{-0.15cm}{$10^5$}}
\psfrag{6.1}{\small \raisebox{-0.15cm}{$10^6$}}
\psfrag{0.2}{\small \hspace{-0.2cm} $0.2$}
\psfrag{0.4}{\small \hspace{-0.2cm} $0.4$}
\psfrag{0.6}{\small \hspace{-0.2cm} $0.6$}
\psfrag{0.8}{\small \hspace{-0.2cm} $0.8$}
\psfrag{1.}{\small \hspace{-0.36cm} $1.0$}
\psfrag{n}{\hspace{-1.0cm} \raisebox{0.1cm}{$2^{-n} \, Z \:, \: 2^{-n} \, Z_n$}}
\psfrag{c1}{$\: T / B$}
\epsfig{file=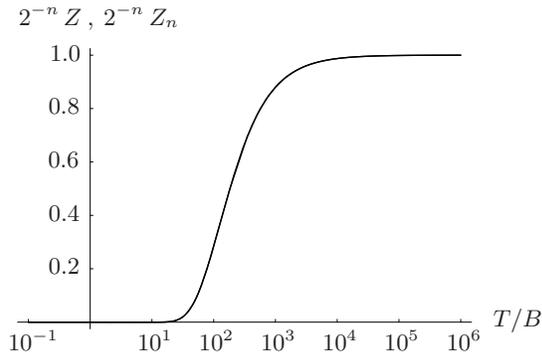,width=7cm}
\caption{Partition function for a chain of 100 spins with $B = K = 1$ divided by the number of states,
$2^{100}$. The difference between
$2^{-n} \, Z_n$ and $2^{-n} \, Z$ is not visible ($Z_n$ and $Z$ are defined in equations
(\ref{partf_def}) and (\ref{partf}) respectively).}
\label{zustf}
\end{figure}

Figure \ref{density_10_1} shows the density of states $\eta_n (E)$ and its approximation $\eta (E)$
for a chain of 10 spins, while figure \ref{density_15_1} shows the same plot for a chain of 15 spins.
The approximation works well despite the still small number of spins; furthermore, the tendency that
the approximation improves with increasing number of spins is evident.
%
%
%
%
%
%
%
Figure \ref{zustf} shows the partition function for a chain of $100$ spins divided
by the number of states, $2^{100}$, as a function of temperature.
The difference between the exact function $Z_n$ and the approximation $Z$ is not visible.
To see whether the convergence improves with the number of spins, we have considered the maximal difference
between $Z_n$ and $Z$ for all temperatures,
\begin{equation}
\delta(n) = \max_T \, \left| Z_n - Z \right|.
\end{equation}
We have found the following values: $\delta(10) \sim 10^{-3}$, $\delta(100) \sim 10^{-8}$ and
$\delta(1000) \sim 10^{-11}$.
%
%
%
%

%
%
\section{discussion and limitations}
\label{discuss}

The approximations of the quantities $\eta_n (E)$ and $Z_n$ with equations (\ref{statedens}) and (\ref{partf})
face some problems that cannot be avoided.

Firstly, the convergence in equation (\ref{limit}) is only weak, i.e. the lhs converges to the rhs
for all intervals $\left[E_1, E_2\right]$ of nonzero length. The convergence is not pointwise.
In the present case this means that, for example, the trace of a projector on a single eigenstate
$P = \ve \varphi \ket \bra \varphi \ve$ cannot be approximated.
The problem becomes apparent, if one tries to calculate the partition function for zero temperature
via (\ref{partf}), where only the ground state is occupied.
In the present model, for example, this state is energetically separated from the other states
that form a quasi continuous band.
As a consequence quantum phase transitions \cite{Sachdev1999} occuring at zero temperature can not be
treated with our approach.

The second drawback of our approach is the following: Each term $\bra a \ve \mathcal{O} \ve a \ket$ for
some operator $\mathcal{O}$ that is a function of $H$
is well approximated and the accuracy increases with the number of subsystems $n$.
On the other hand, the number of terms in the sums (\ref{statedens}) and (\ref{partf}) increases
exponentially with the number of subsystems, for example with $2^n$ for the spin chain. 
Therefore, the quantities $\eta (E)$ and $Z$ can only be in good accordance with
$\eta_n (E)$ and $Z_n$, if both are divided by the dimension of the Hilbert space, i.e. the number of
states $\ve a \ket$.

This will not always be problematic, since the errors each term $\bra a \ve \mathcal{O} \ve a \ket$
carries need not all be of the same sign and may thus cancel each other, as in the calculation
of the spectral densities.

In the calculation of the partition sum, however, there is the following problem:
Every stable system has a finite minimal energy, the energy of the ground state. Nonetheless
the probability density (\ref{prob_dens}) is, albeit very small, nonzero for all energies.
Therefore, one needs to introduce the cutoffs in the integral (\ref{expect_O}). As can be seen from
expression (\ref{partf}), the upper limit of the integral does not matter.
The lower limit on the other hand matters and becomes increasingly relevant at low temperatures.
No matter what lower limit of the integral we take, the error of the approximation (\ref{partf})
always has the same sign. 

%
%
%
\begin{figure}
\bigskip

\psfrag{-3.1}{\small \raisebox{-0.15cm}{$10^{-3}$}}
\psfrag{-2.1}{\small \raisebox{-0.15cm}{$10^{-2}$}}
\psfrag{-1.1}{\small \raisebox{-0.15cm}{$10^{-1}$}}
\psfrag{1.1}{\small \raisebox{-0.15cm}{$10^1$}}
\psfrag{2.1}{\small \raisebox{-0.15cm}{$10^2$}}
\psfrag{3.1}{\small \raisebox{-0.15cm}{$10^3$}}
\psfrag{4.1}{\small \raisebox{-0.15cm}{$10^4$}}
\psfrag{0.1}{\small \hspace{-0.2cm} $0.1$}
\psfrag{0.2}{\small \hspace{-0.2cm} $0.2$}
\psfrag{0.3}{\small \hspace{-0.2cm} $0.3$}
\psfrag{0.4}{\small \hspace{-0.2cm} $0.4$}
\psfrag{0.5}{\small \hspace{-0.2cm} $0.5$}
\psfrag{0.6}{\small \hspace{-0.2cm} $0.6$}
\psfrag{0.7}{\small \hspace{-0.2cm} $0.7$}
\psfrag{n}{\hspace{-1.3cm} \raisebox{0.1cm}{$\ln(Z)/n \:, \: \ln(Z_n)/n$}}
\psfrag{c1}{$\: T / B$}
\epsfig{file=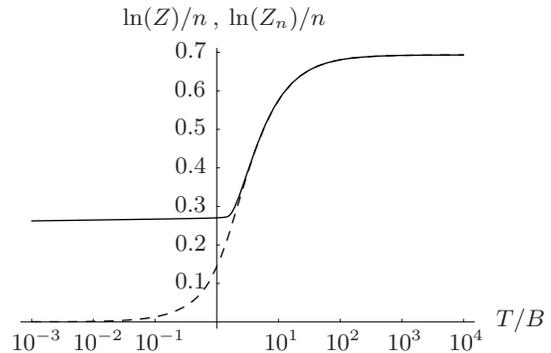,width=7cm}
\caption{Logarithm of the partition function divided by the number of spins for a chain of 1000 spins with
$B = K = 1$. The dashed line shows the exact expression
$\ln(Z_n)$ and the solid line the approximation $\ln(Z)$ ($Z_n$ and $Z$ are defined in equations
(\ref{partf_def}) and (\ref{partf}) respectively).}
\label{zustflog}
\end{figure}
Figure \ref{zustflog} shows the logarithm of the exact partition function of the
spin chain and its approximation, each divided by the number of spins, for a chain of 1000 spins.
The plot is done with the exact ground state energy. For low temperatures ($T < B$), the value of the
approximation is too large and the approximation fails.
Therefore, only the partition sum divided by the number of states can be accurately predicted.

\section{summary}
\label{summary}

We have given analytic expressions for spectral densities and partition sums of chains of
quantum systems with nearest neighbor interaction, which are valid in the limit of infinitely many
subsystems. We have numerically evaluated these expressions for a spin chain with a finite number of spins
and compared the results with the values obtained by exact diagonalization. The results show increasing
accordance with growing number of spins.

Furthermore, we have discussed the limits of the validity of our approach and
some problems that can occur, as well as their effects.

The results of this paper should provide useful tools for the calculation of spectral
and thermodynamical quantities in many systems which are intensively studied in present day
condensed matter physics. Among those are spin chains \cite{Wang2002} and strongly correlated electrons
\cite{Korepin1994}.

We thank M.\ Henrich, C.\ Kostoglou, M.\ Michel, H.\ Schmidt, M.\ Stollsteimer and F.\ Tonner
for fruitful discussions.

M.H., in particular, wants to thank Prof. Detlef D\"urr for intensive discussions and many helpful comments.

%
%

\end{document}